# A Service-oriented Infrastructure Approach for Mutual Assistance Communities

Ning Gui, Hong Sun, Vincenzo De Florio, Chris Blondia

*University of Antwerp*
*Department of Mathematics and Computer Science*
*Performance Analysis of Telecommunication Systems group, Antwerp, Belgium*
*And Interdisciplinary institute for BroadBand Technology, Ghent-Ledeberg, Belgium*
*{ ning.gui, hong.sun, vincenzo.deflorio, chris.blondia}@ua.ac.be*

## ABSTRACT

Elder people are becoming a predominant aspect of our societies. As such, solutions both efficacious and cost-effective need to be sought.

This paper proposes a service-oriented infrastructure approach to this problem. We propose an open and integrated service infrastructure to orchestrate the available resources (smart devices, professional carers, informal carers) to help elder or disabled people. Main characteristic of our design is the explicitly support of dynamically available service providers such as informal carers. By modeling the service description as Semantic Web Services, the service request can automatically be discovered, reasoned about and mapped onto the pool of heterogeneous service providers. We expect our approach to be able to efficiently utilize the available service resources, enrich the service options, and best match the requirements of the requesters.

## 1. INTRODUCTION

As well known, the proportion of elderly people keeps increasing since the end of last century. The European overview report of Ambient Assisted Living (AAL) investigated this trend [1]. The studies of EUROSTAT [2] indicated that the share of the total European population (EU 15) older than 65 is set to increase from 16.3% in 2000 to 22% by 2025 and 27.5% by 2050, while that over 80 (3.6% in 2000) is expected to reach 6% by 2025 and 10% by 2050.

Studies of Counsel and Care in UK found out that these elderly people would prefer to live in their own house rather than in hospitals, thus they need support to remain independent at their home [3]. Researches also showed that living home could save the caring expense: for instance, an Alzheimer patient will cost $ 64,000 in nursing home while only $ 20,000 at home [4]. The problem raised by living home is that people's daily lives are hampered by aging [2] (see Fig. 1). In order to improve the quality of life for the elderly and disabled people, it is important to guarantee that assistance to those people be timely arranged in case of need.

 "Big projects" are launched to assist the elderly people living independently in their own house: in the European Union, Article 169 (of the EC treaty) states that the European Commission may allocate a 700M€ budget to AAL [5].

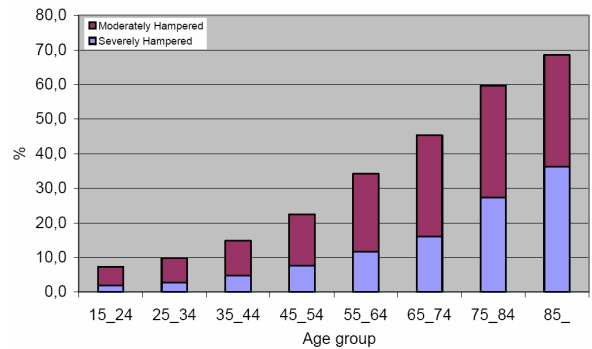

Fig. 1 Percentage of people hampered in daily activities, by age[2]

The main contributions of this paper can be summarized as follows:

Under these AAL projects, assistive devices are developed to facilitate the daily lives of these elderly and disabled people. These devices help the elderly and disabled people to overcome their handicaps and improve their independency, but they cannot constitute the sole ingredient for an effective solution to the AAL challenges: For instance, in the AAL country report of Finland it was remarked how "individual assistive devices have shown the potential of such products in improving independence and quality of life of the users". [6]. However, in the same report, it is also pointed out that those "devices are not useful if not combined with services and formal or informal support and help". We share this view and deem informal carers as indispensable when constructing timely and cost-effectively services to assist the elderly people. Further evidence to this claim comes from our previous research, where we developed a design tool to evaluate the performance of informal carers in so-called mutual assistance communities [7], i.e. communities whose members may request assistance and at the same time get motivated to play the role of caregivers. Simulations have shown that informal carers are capable to contribute effectively to the community welfare.

The proposed system, that we call iMACom for service-oriented Infrastructure for Mutual Assistance Communities, mediates among all the available resources with the goal of providing timely and cost-effective service to the elderly people. This ambitious goal comes in particular from expanding the role of

service providers, which raises in particular two research challenges:

1. Firstly, the assumption that the service in a mutual assistance community is dynamic means that a service provider may join, update or leave at any time. How to cope with this dynamicity becomes a challenge.
2. Secondly, the service provider is heterogeneous in nature (computer, human, and hybrid) and capabilities; a service may be deployed in various forms. As a consequence, how to define and map request to service, and how to reflect the preference of requester become challenges.

Our system is designed to meet these challenges by incorporating a service oriented infrastructure. The concept of service oriented architecture was designed to explicitly support dynamic availability of services in nature. At the same time, a semantic analysis mechanism is employed in our infrastructure to reason and map the service request to the heterogeneous service provider.

The remainder of the paper is organized as follows: In Section 2, related work will be reviewed. In Section 3, an overview of the mutual assistance community is presented. The service architecture of the iMACom system is stated in Section 4. Section 5 gives scenarios of the operation of our iMACom system. Conclusions and future work are given in Section 6.

In this section, we summarize related work focusing on two aspects related to our work – research on assistive systems design through Ambient Intelligence, and research on conceptual tools for service discovery and mapping.

## 2. Related work

In this section, we summarize related work focusing on two aspects– research on assistive systems design through Ambient Intelligence, and research on conceptual tools for service discovery and mapping.

### 2.1 AAL design

Numerous researches are being carried out on building intelligent environments around people, such as such as Aware Home [8], Smart In-Home [9] BelAmI [10], I-Living [11]. The common feature of these researches is to deploy wireless devices to detect the status of the occupant, collect these data by an intelligent terminal, and trigger consequent actions. Devices such as RFID, motion or fall detectors, etc. are used to accomplish tasks such as activity reminding, health monitoring, personal belonging localization, emergency detection, and so forth.

These researches on "smart house" improved the independency of the elderly people, and reduced the required manual work. However, the limitation for the "smart house" approach is that the researches are mainly focused inside the house where the elderly people reside; few attention is paid on the communications with the outside world, mainly restricted to emergency calls and occasional contacts with the family. Negative effects are as follows: firstly, the devices need to work coherently with people to reach its potential. Secondly, the lack of communication with the community outside the house inherently limits the service exploration, and may isolate the user from the outside world.

Other projects moved the focus on the communication with the outside community. One such project is COPLINTHO [12], which built an eHomeCare system combing forces from the patient's family, friends and overall care team. The limitation of this class of investigations is that the application is restricted to the recovery progress of a patient, thus the communications between different players are mainly focused on exchanging the medical data of the patient. In this view, the service model in these projects is not sophisticate enough to cope with the challenges of a service-oriented infrastructure for AAL.

Finally, some efforts aim to assist the aging people by building a community. One such case could be found in Beer's work [13] [14] [15]. Beer developed an INCA (Integrated Community Care) system with AUML (Agent-oriented Unified Modeling Language), intending to develop care plans for the elderly people by coordination [14]. Informal carers are making contributions by providing non-medical carer. Beer's system is the one more similar to the one described herein, with several important limitations, including the absence of the mechanism to locate and map services, and only scheduled the professional carer currently.

All these above represent examples of the existing approaches to AAL, which we may classify as person-oriented, family-oriented, and community-oriented. Their achievements have both inspired our work and provided useful provisions, methods, and architectures that integrate our contribution. None of these methods is a true combination of technical and social forces, like we envisage in our infrastructure. Moreover, none of them truly defines a service-oriented infrastructure for the orchestration of AAL services, where human assets and computer resources are both considered and exploited. Our system is constructed to coordinate optimally the demand and the offer for services. Semantic reasoning is adopted for request analysis and service mapping.

### 2.2 Service discovery and mapping mechanisms

As already mentioned, in mutual assistance communities, service requesters and providers, even from the same ontology domain, are heterogeneous in nature and capabilities. This heterogeneity may jeopardize the effectiveness of any infrastructural approach to AAL, as it hampers the effectiveness of service discovery and mapping between the request and service.

Industry efforts to standardize web service description, discovery and invocation have come to standards such as WSDL [28] and UDDI [27]. However, these standards, in their current form, suffer from the lack of semantic representation. Richer semantic specifications of services are needed to cope with this challenge. Realizing this need, the semantic web community proposed to formalize the semantic representations with the help of model theory and description logics, producing respectively the RDF [29] and OWL [17] languages.

Several approaches have been proposed to semantically represent Web services. One such approach aims at developing new semantic web based standards for web services [30]. A second approach - METEOR-S[31] adds semantics to existing Web services standards.

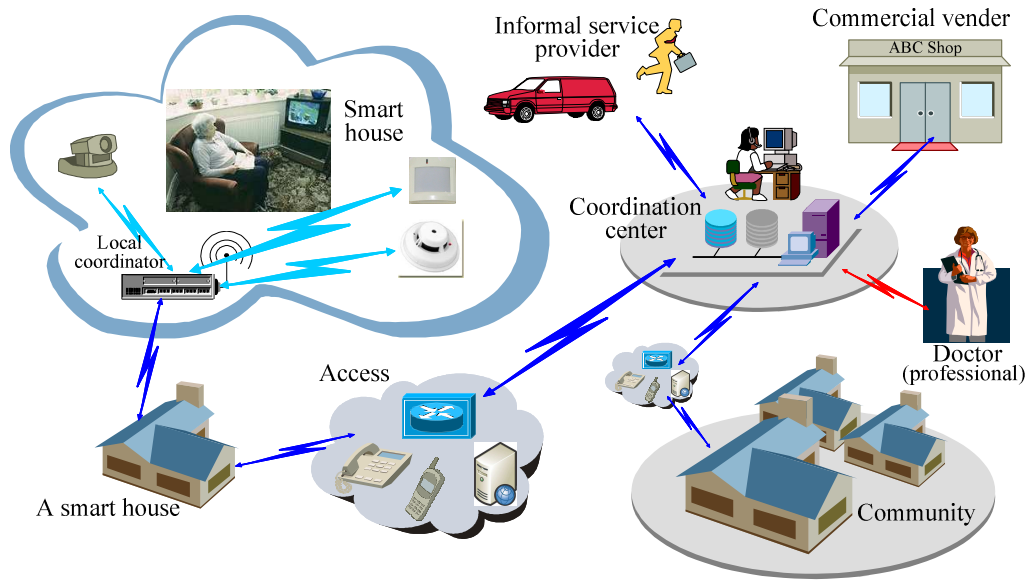

Figure 2: Mutual Assistance Community organization

In our infrastructure, we modeled the service description as Semantic Web Services that can automatically be discovered and accessed by semantic match service. Services are matched based on the ontological classifications of their semantic service descriptions.

## 3. Key ingredients in iMACom

Our work takes inspiration from many of the ideas in these projects but recasts them in the light of service-oriented infrastructures. Much work has been done in Ambient Assisted Living either in the technology domain or in the societal domain, but few of them, to the best of our knowledge, have focused on an infrastructural design combining both fields.

By providing uniform abstractions and reliable services for common operations, service infrastructures could make it easier to develop and incorporate a diverse and constantly changing set of service providers, human or devices. A service infrastructure would also make it easier to incrementally deploy new sensors, new devices, and new services as they appear in the future, as well as scale these up to serve large numbers of people. Lastly, a service infrastructure would help people and devices to share their requests and service data, placing the burden of acquisition, processing, and interoperability on the infrastructure instead of on individual devices and service. These advantages brought us to employing an infrastructure approach to AAL.

This infrastructure aims at fostering a mutual assistance community. By integrating the current research on hardware such as smart housing and software technologies and, the most important thing, we think, the people of the community, it allows disparate technologies and people working together to helping people who suffer from aging or disabilities. Some key ingredients of iMACom are analyzed in the following sections.

### 3.1. Service Requester

In the mutual assistance community, a requester entity is a person or smart device that wishes to make use of a provider entity's service. The service requester (SR) can be people who either explicitly required one type of service or an intelligent device which raises a service request by reasoning on the context-data provided by, e.g., sensors.

For instance, when the smart device monitoring through some BAN sensor detects that the blood pressure (or other health-relevant figures) is dangerously high, or through some visual sensor that the person has been immobile on the floor for an unreasonably long time, an emergency health care request is raised to the infrastructure.

Normally, the service requester raises their service request by specifying requirements such as service type, scheduling time, requester authentication, estimated duration, priority, location,, etc. After the procession of the Format Service(described in section 4.1), the request will be described in an ontology based semantic description to facilitate later semantic service matching . Some user-interface and the semantic description languages should be designed and employed to express the service request. The ongoing Protégé project [21], which is developing a suite of tools to construct domain models and knowledge-based applications with ontologies, may help to mitigate this problem.

### 3.2. Service Provider

As mentioned already, our approach involves both professional and non-professional people as possible service providers (SP). A service provider can also be a smart device providing information or actuating actions. In the rest of this section we describe these main classes of service providers.

**Professional Service providers:** these are those who have specialized intellectual or creative expertise based on personal skills. This type of service providers consists of professionals, private firms, professional organizations or municipalities. Normally, the type of service may have better quality and higher availability but the cost of this service is comparatively higher. The doctor and commercial vender in Fig. 2 are two examples of Professional service providers.

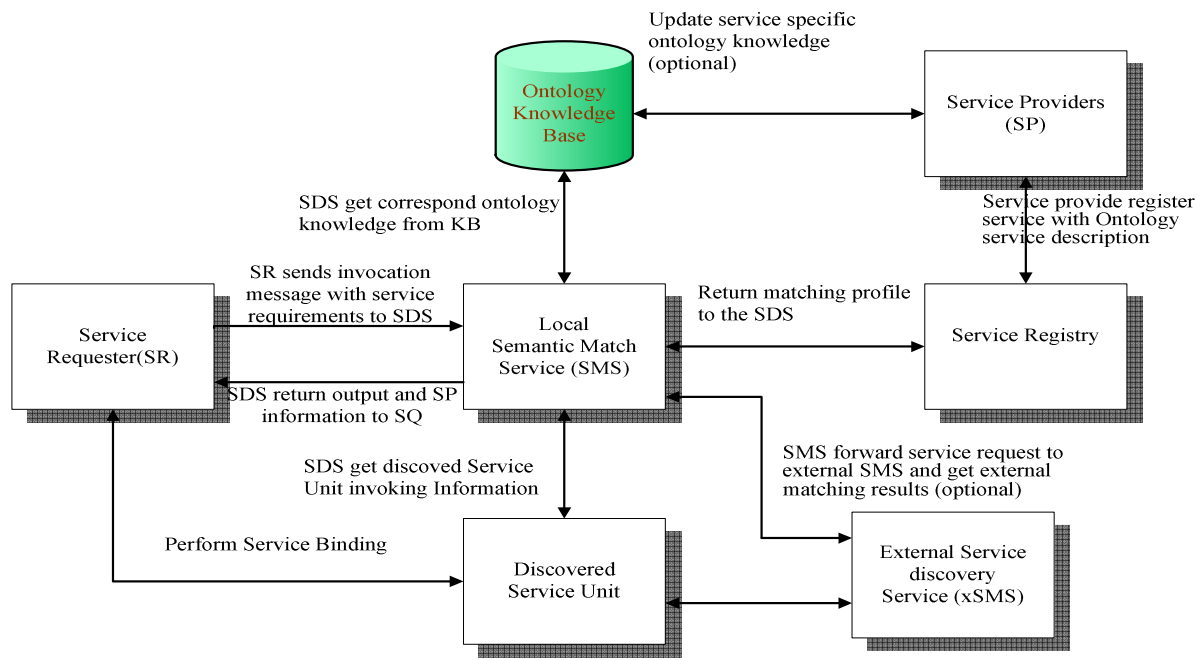

Figure 3: The iMACom Architecture

**Non-professional service providers**: also known as "informal carers". This type of service providers normally is made of ordinary people, volunteers or specifically employed people. Specialized expertise is not necessary for providing this type of service. Some of these people, under certain circumstance, will register to do some non-professional work such as gardening, housing or transportation either volunteer or for income. In some cases, service providers with proper training can provide some professional work such as first-time aid when the professional people are unavailable. The characteristic of their service is the dynamicity of their availability.

**Smart Device**: smart device technologies can provide some sort of service for the assisted people (AP). Normally, these service are located in the smart house domain. These devices can provide services such as health monitoring, alerting, personal belonging localization etc. With the help of the smart device, some task can be performed to ease the AP's life.

### 3.3. Coordination center

Every community has a coordination center, which is designed to coordinate the services and requests inside this community and establish connections with other communities. In a community, the coordination center (CC) deals with: providing service registration for the service provider; storing service information; accepting the user request and mapping it to the best fitting service. In iMACom, each community has a certain boundary due to technology and social restrictions, however, the coordination center of each community could be linked together, share their services and exchange information, thus breaking through the boundaries to organize a larger community. In Fig. 2, a "smart house" could be considered as a small community under the control of a local coordinator center. Connecting the local coordinators together we come up to the overall community in Fig 2. Moreover, the coordination center in Fig. 2 could also connect with coordination centers in other communities to set up a community in larger scale. How the coordination center works is introduced in details in section 4.

### 3.4. Interactions via iMACom

In the house domain, we envisage an assisted living supportive environment has at least one local intelligence node which acts as local CC in the house domain, this house CC can be a specialized PC, PDA, or a black box equipped with one or more wireless interface cards (IEEE 802.11, Bluetooth, Ultra Wide Band, or Infrared). In the service-oriented model, each smart device should register its service capacities in the house CC with its service descriptions, such as the Smart TV can register "TV", "Text display" service and etc. The smart phone can register for the "Phone service", "Context display" and "Network connection" service via GPRS or 3G. In the community domain, each people or organization will register their service in the community CC with the semantic information about service type, availability, prices and etc.

From figure two, we can see the doctor and the commercial vender will register in the community CC to provide service. The non-professional people also register their capability in this center. For instance, the hospital may register for the "emergency medical care" service with availability of 24 hour per day and a volunteer may register for the "gardening" for limited periods of time. The CCs are deployed hierarchically, as

When a request is raised either by the AP or smart device, the house CC will first perform local semantically match in house domain service registry. If any match achieve, it will help binding this service provider with requester. Service was invoked and the request will be fulfilled. If the ALH could not find a match, it may forward this request to the community server. The server will run the semantic match in the community service registry, if match, the following step will be the same as in the house domain. Otherwise, it will return a no matching result to the house CC.

## 4. The iMACom Architectures

Based on our previous analysis, we design the iMACom architecture which aims to provide an efficient infrastructure support for building AAL community. It consists of the following infrastructure service which acts as basic service components.

### 4.1. Format Service

This service helps to collect request and service descriptions from heterogeneous sources – people or devices – , and convert them to semantic representations so that request information can be shared and reasoned upon by other service components.

Normally, the SR requires a service for the functional aspect. Due to the great variety of user's habits, environments, income etc. and heterogeneous service provider, the format service should be able to represent the semantics of both the functional and non-functional requirements in their descriptions. For instance, in [25] profiles are used to integrate the non-functional requirements in the service requests. We deem that both SR and SP should also state the non-functional restrictions in their descriptions, such as price, quality, service provider type etc. For instance, user can chose the best service quality as preference which normally means he or she should pay more for a high quality service, or the other one may have preference on cheapest SP.

For the user, the Format Service should provide a light-weight, easy-to-use interface. This interface should be easy-to-use, safe, adaptive with respect to user mistakes. For the smart device, due to the resource limitation of pervasive devices such as CPU speed and memory, it may not be able to implement the format service by itself. A smart device with plenty of resources may act as a formatting broker and help this device to perform formatting task.

### 4.2. Service Registry

Service provider registers their services in directories along with profiles that describe their various relevant capabilities and characteristics. Most service registries are ased on the UDDI standard, which focuses on registration of service descriptions. UDDI provides information about the entity that owns it and provides mechanisms to classify the service in terms of standard taxonomies such as North American Industry Classification System (NAICS) [9].

In order to perform the semantic discovery service, service registry should be extended to contain semantic service descriptions. The service registry module needs to translate the ontologies based descriptions to UDDI records and then publish the given services under the specified taxonomy (such as NAICS) in UDDI. Reference [12] prescribed a mapping scheme to perform this task.

### 4.3. Semantic Match Service

The Semantic Match Service (SMS) is responsible for semantic processing, reasoning and matching the request to service by employing logic reasoning. We use a demonstration first to show why this service is critical in our iMACom infrastructure.

SR **A** raised a "watering flower" request for feeding his flowers. If SP **B** provides the same type of service, the SDS will return the matching service provider. But things may get more difficult if **B** registered for "gardening" service instead of "watering flower" service. Comparing these two services only through syntactical metrics will produce a negative response. This leads to false negatives. Because, the "watering flowers" is a part of "gardening" service, normally, **B** should able to provide the watering service.

This demonstration shows that lack of semantics hampers the effectiveness of computing the compatibility between the request and service. In order to perform semantics based match, an ontology-based approach should be used to describe service semantically. This model should able to enable formal analysis of domain knowledge in a way which is independent of programming language, underlying operating system or middleware. A good candidate for this model is the Web Ontology Language (OWL). The SMS matches functional and non-functional requirements of a service request with a service provider's ones. Two properties are considered a match if they either match exactly, or as defined by some relationship that can be inferred from the ontology using an inferring engine which employs the ontology relationship and the rules in the Knowledge Base. By specifying this type of matching criteria, this allows for matches that are close though not exactly equivalent. In the above example, the match results from the "watering" service holding an 'Is-a' relationship with 'Gardening' in the Gardening domain ontology.

Normally, this service consists of a semantic service reasoner and an ontology-based Knowledge Base (KB).

#### 1.1.1 Semantic Service Reasoner

The semantic service reasoner has the functionality of providing deduced ontology information from the ontology service descriptions and the KB. Multiple logic reasoners can be incorporated into the semantic service reasoner to support various kinds of reasoning tasks. Different inference rules can be specified and preload into various reasoners. Developers can easily create their own rules based on predefined format.

#### 1.1.2 Ontology-based Knowledge Base

Knowledge base (KB) consists of a set of sub-domain ontologies in the different domains(housing, transportation, nursing, etc.) and a set of correspond rules in these sub-domains. These ontologies describe the concepts and relationships from the application domain. The rules are the problem solving procedures expressed with the terms from the ontology. The ontology information can be manually predefined or can be updated by dynamically parsing the ontology-based profiles of service descriptions. The KB also should provide a set of API's for other service components to query, add, delete or modify the KB.

The SMS can be connected together, when no matching service could be found in a local service registry. It can forward the service request to the external SMS. The external SMS will perform semantic Match on its local service registry. This scheme facilitates SMS and, in the meanwhile still is able to use all possible resources.

### 4.4. Scheduling service

Another challenge lies with the effectively usage with the resources in the community while satisfies different task's requirements.

Sometimes, there are several services which meet the service request, how to select the appropriate one from the candidates will depend on many criteria. For example, there are several medical professionals attached to the community who provide similar medical services. The scheduler may arrange the request to the professional which is closer to the patient to save the traffic load. In the schedule service, the higher priority service request can be scheduled to preempt low priority tasks. A doctor may be required to deal with nearby emergency medical events as soon as possible rather than continue to perform routine medical check for an elderly people.

### 4.5. Service Binding

After the SMS and scheduling Service,, If suitable service providers are present in the service registry, the binding service will provide a scheme to sign a "contract" between the service requester and service provider. This contract can ensure the requester gets required service and the service provider gets the correspondent return the "contract" can be exploitted in late accounting & auditing usage. This service also provides the participants additional information about this transaction such as correspond methods and etc.While service binding process of the service which provided by smart device or software, this process can be map to a simple process - returns a reference to a service object that implements the service functionality and the ways to invoke this service object.

## 5. Scenario Analysis

We envisage there are several device and companies register their services in local service registry the house domain or in the community service registry.

**In house domain:**
Each of the smart devices such as the sensors, actuators, displays should register their capability to local service register as described in section 3.4.

**In community domain:**
The people or organizations who want to provide service will be required to register in the community service registry server. Services are registered in directories along with profiles that describe their various relevant capabilities and characteristics. In this scenario, we envisage the telephone company to register in the community service registry for a "reminder" service.

**Remind Service Scenario**
Remind service is a service that helps the AP to follow his/her cyclical routines, such as taking medicine or taking vital signs.

**Tradition AAL approach**: when it is time for the AP to carry out his/her routines the local CC will invoke the pre-assigned devices such as TVs, cell phones, wearable headsets, and send reminder messages to these devices [12]. In this approach, the requester & service provider are tightly coupled. The static service mode could not deal with situations when service are dynamically and automatically added or removed from the system. This type of approach makes the AAL system fragile and rigid.

**iMACom infrastructure based Approach:**
In our framework, service is not confined to a particular provider; instead, service providers are substitutable as long as they semantically comply with the same service description, service provider can be smart devices or professional or non-professional people. Another fundamental characteristic of service orientation is that services exhibit dynamic availability, since they can be added or removed from the service registry at any time. We demonstrate the reminder service in two different situations.

**With smart device**: the request is sent directly to the local CC with service restrictions. The local CC will perform semantic service discovery in the local service registry first. When it perform service match, it will not find the match service by only employ syntactic matching. By semantic inference, the "Text display" service can be matched to the "Reminder" service request because the reminder service can be produced by displaying a reminder sound. The smart TV and the smart phone will be the good candidate for the remainder service request after semantic match. The local CC will select one of the two by user preference or by context-awareness.

**Without smart devices at home**: When the house is not equipped with smart TV or smart phones, the local CC could not find a match service provider. In this case it will raise its request to the community service discovery service. The latter will perform semantic service discovery in the community domain. Now, the telephone company will be matched because it registers the "Remind service". So it will help the service request to set a contract with the telephone company and the telephone will provide the reminder service. Of course, some volunteers or professional service providers would also provide "reminder" services. If available, the community may also select one from this candidate through user's preference or other factors

## 6. Conclusion

The need for the assisted living supportive environment is compelling, due to the increasing economic and social problems posed by aging. In this paper, we provided an overview of iMACom, a service-oriented infrastructure for mutual assisted community aimed at helping assisted person's daily life by orchestrate all the possible resources. In this infrastructure, the smart device, non-professional people and professional people can dynamically register / unregister their service in the service registry. In the service processing aspect, different types of service provider and different services are unified together and treated equally to provide comprehensive and all-level service to the AP. The AP can get help from all available resources, e.g. smart device, volunteers, private firms or the municipality, etc, while at the same time the societal costs of the corresponding resources can be contained.

Within iMACom, services are described with ontology description and represented as Semantic Web Service. SMS is designed to perform semantic match. SMS.determines if a service is suitable for a particular request by semantically matching both the functional and non-functional capabilities and requirements of a service are matched with those of a request. The SMS connects with external SMS which enables them to forward internal service request and discovered outer services which match the requirements.

In the near future we shall design a prototypic iMACom and exercise it in simulated and real-life situations to validate its ability to fully comply to our design goals